\documentclass[prd, unsortedaddress, superscriptaddress, twocolumn, twocolappendix,nobibnotes,reprint,nofootinbib]{revtex4-2}
\usepackage[T1]{fontenc}
\usepackage{lmodern}
\usepackage{amsmath}
\usepackage{bm}
\usepackage{tensor}
\usepackage{booktabs}
\usepackage{graphicx}
\usepackage{xcolor}
\usepackage{amssymb}
\usepackage{hyperref}
\usepackage{aas}
\usepackage{orcidlink}
\usepackage{mathtools}

\hypersetup{
    colorlinks=true,   
    linkcolor=blue,    
    filecolor=blue,    
    citecolor=blue,    
    urlcolor=blue,     
    pdfborder={0 0 0}  
}

\usepackage[nameinlink,noabbrev]{cleveref}
\crefname{equation}{Eq.}{Eqs.}
\Crefname{section}{Section}{Sections}
\Crefname{table}{Table}{Tables}

\renewcommand{\vec}[1]{\mathbf{#1}}
\renewcommand{\d}[0]{\mathrm{d}}

\newcommand{\uhm}{Department of Physics and Astronomy, University of Hawai`i at M\=anoa, 2505 Correa Rd., Honolulu, HI 96822 USA}
\newcommand{\asusese}{School of Earth and Space Exploration, Arizona State University, Tempe, AZ 85287 USA}
\newcommand{\asuphys}{Department of Physics, Arizona State University, Tempe, AZ 85281 USA}
\newcommand{\bu}{Department of Physics, Boston University, 590 Commonwealth Avenue, Boston, MA 02215 USA}
\newcommand{\iac}{Instituto Avanzado de Cosmolog\'{\i}a A.~C., San Marcos 11 - Atenas 202. Magdalena Contreras. Ciudad de M\'{e}xico C.~P.~10720, M\'{e}xico}

\begin{document}

\author{Kevin~S.~Croker\orcidlink{0000-0002-6917-0214}}
\email{kevin.croker@asu.edu}
\affiliation{\asusese}
\affiliation{\uhm}

\author{Joshua~Cohon\orcidlink{0009-0004-4799-891X}}
\affiliation{\asusese}

\author{Viktor~T.~Toth\orcidlink{0000-0003-3651-9843}}
\affiliation{Ottawa, K1N 9H5 Ontario, Canada}

\author{Massimiliano Rinaldi\orcidlink{0000-0003-0325-3911}} 
\affiliation{Department of Physics, University of Trento, Via Sommarive 14, 38122 Trento, Italy} 
\affiliation{Trento Institute for Fundamental Physics and Applications TIFPA-INFN, Via Sommarive 14, 38122 Trento, Italy} 

\author{Nicolas~Fernandez\orcidlink{0000-0002-3573-339X}}
\affiliation{NHETC, Department of Physics and Astronomy, Rutgers University, Piscataway, New Jersey 08854 USA}

\author{Joel~L.~Weiner\orcidlink{0000-0002-1888-8744}}
\address{Department of Mathematics, University of Hawai`i at M\=anoa, 2565 McCarthy Mall, Honolulu, HI 96822, USA}

\author{Duncan~Farrah\orcidlink{0000-0003-1748-2010}}
\affiliation{\uhm}
\affiliation{Institute for Astronomy, University of Hawaii,  2680 Woodlawn Drive, Honolulu, Hawaii, 96822 USA}

\author{Gregory~Tarl\'e\orcidlink{0000-0003-1704-0781}}
\affiliation{Department of Physics, University of Michigan, 450 Church Street, Ann Arbor, Michigan 48109, USA}

\author{Gustavo~Niz\orcidlink{0000-0002-1544-8946}}
\affiliation{Departamento de F\'{\i}sica, DCI-Campus Le\'{o}n, Universidad de Guanajuato, Loma del Bosque 103, Le\'{o}n, Guanajuato C.~P.~37150, M\'{e}xico}
\affiliation{\iac}

\author{Christopher Cain\orcidlink{0000-0001-9420-7384}}
\affiliation{\asusese}

\author{James~W.~Rohlf\orcidlink{0000-0001-6423-9799}}
\affiliation{\bu}

\author{Alejandro~Aviles\orcidlink{0000-0001-5998-3986}}
\affiliation{Instituto de Ciencias F\'{\i}sicas, Universidad Nacional Aut\'onoma de M\'exico, Avenida Universidad s/n, Cuernavaca, Morelos, C.~P.~62210, M\'exico}
\affiliation{\iac}

\author{Jake~Summers\orcidlink{0000-0002-7265-7920}}
\affiliation{TAPIR, California Institute of Technology, Pasadena, CA 91125, USA}
\affiliation{LIGO Laboratory, California Institute of Technology, Pasadena, CA 91125, USA}

\author{Leander~Thiele\orcidlink{0000-0003-2911-9163}}
\affiliation{Center for Data-Driven Discovery, Kavli IPMU (WPI), UTIAS, The University of Tokyo, Kashiwa, Chiba 277-8583, Japan}
\affiliation{Kavli IPMU (WPI), UTIAS, The University of Tokyo, 5-1-5 Kashiwanoha, Kashiwa, Chiba 277-8583, Japan}

\author{Rogier~A.~Windhorst\orcidlink{0000-0001-8156-6281}}
\affiliation{\asusese}

\author{Joshua~Grumski-Flores\orcidlink{0009-0004-2634-5223}}
\affiliation{\asuphys}

\title{A Strong Cosmologically Coupled Mass Growth Solution in Linearized General Relativity}
\begin{abstract}
  DESI Data Release 2 has heightened interest in cosmologically coupled black holes (CCBHs) as an explanation for accelerating late-time expansion, but known models for CCBHs grow too slowly to drive it. 
  We construct at linear order a solution to General Relativity whose Hernandez-Misner-Sharp mass passively grows $\propto a^\kappa$ as the universe expands.
  The constant $\kappa$ enters all eigenvalues of the Weyl and stress-tensor operators, so growth is unambiguously physical in both free-field and material sectors.
  The solution is causal over spacetime relevant for stellar collapse compact objects.
  A cloud of energized vacuum emerges at $\kappa = 3$, complementing the $a^3$ scaling predicted elsewhere for the masses of energized vacuum black holes.
  This first step toward $\kappa \sim 3$ CCBH exact solutions enables observational tests with pulsar binaries.
\end{abstract}

\maketitle

\section{Introduction}
Cosmological coupling is a recent proposal in general relativity (GR) that a global Robertson-Walker (RW) cosmology and local regions of relativistic stress within it can strongly couple \citep{CrokerWeiner19}.
The utility of this coupling occurs in combination with long-studied energized vacuum \citep{Gliner66} non-singular black hole (BH) solutions in GR%
\footnote{Non-singular BH solutions roughly classify as (i) Frolov, Markov, Mukhanov (limiting-curvature hypothesis) — spacelike junction inside the horizon using an exotic surface layer and therefore Israel’s conditions, (ii) Gravastars (e.g. Mazur–Mottola) — the surface layer is outside the horizon, (iii) Smooth-interpolation ``regular black holes'' (e.g. Dymnikova, Hayward, Bardeen), and (iv) topology change below the horizon (e.g. Conboy-Lake).}  \citep[e.g.][]{Bardeen1968,Dymnikova92,Dymnkova02,VisserWiltshire04,ConboyLake05,CattoenFaber05,Lobo06,MazurMottola15}.
A population of such BHs can then grow in mass $\propto a^3$, while diluting in number density $\propto 1/a^3$.
In aggregate, they contribute an approximately constant dark energy (DE) density today of the observed magnitude without fine-tuning, because BHs are produced in the deaths of massive stars at $z \lesssim 30$ \citep[c.f.][]{Afshordi08,PrescodAfshordi09}.
The first observational indication for this scenario appeared in supermassive BH (SMBH) growth $\propto a^{3.11}$ within quiescent elliptical galaxy populations at $z \lesssim 2$ \cite{Farrah23a,Farrah23b,Farrah25}.
Recent studies of DESI baryon acoustic oscillation data find complementary support: a time-evolving DE density \cite{DESIDR2BAO} that tracks star formation \cite{CrokerTarle24} is consistent with the cosmic microwave background and gives an expansion rate today $H_0$ closer to local Cepheid-calibrated measurements \citep{RiessYuan22,RiessScolnic24}.
Unlike $\Lambda$CDM and $w_0w_a$ \cite{desi-neutrino-dr2}, the scenario recovers a positive summed neutrino mass \cite{Ahlen25} consistent with terrestrial oscillation measurements \cite{Esteban24} even with low $\ell$ EE CMB measurements and Lyman-$\alpha$ forest-informed constraints on the reionization optical depth $\tau_{\rm rei}$ \cite{CainEngelen25,Elbers25,Kageura26}.

While the approach of \citet{CrokerWeiner19} makes clear statements about cosmological dynamics in the presence of relativistic stress, their global methods cannot constrain physics at kilometer scales where relativistic stress may appear during stellar gravitational collapse.
Thus, complementary insight must be sought through the logically independent investigation of local solutions with RW asymptotics.
Although \citet{FaraoniRinaldi24} have recently established that BH horizons cannot remain static in a RW spacetime and indeed approach comovement \cite{FaraoniRinaldi26}, a clear physical picture of cosmological coupling remains elusive and the subject of much debate~\cite{McVittie33, BlauGuendelman87, FaraoniGao09, CarreraGiulini10, KaloperKleban10, LakeAbdelqader11, GoswamiEllis12, NandraLasenby12b, CrokerWeiner22, Mistele23, WangWangARXIV, Avelino23, Parnovsky24, GaurVisser24, DahalMaharana24, JayswalKopeikin25, FaraoniJacques07, CadoniSanna23, CadoniMurgia24, CadoniPitzalis24, CadoniPitzalis25, CadonideLima26, WuChu26, CalzaPedrotti26,Hayashi26}.
Because existence does not imply uniqueness in GR \cite{BenderOrszag78}, enumeration of decoupled solutions \cite{WangWangARXIV, Avelino23, Parnovsky24, GaurVisser24, DahalMaharana24, JayswalKopeikin25} does not exclude solutions that couple, and construction of coupled solutions \citep[e.g.][]{FaraoniJacques07,CadoniSanna23,CadoniMurgia24, CadoniPitzalis24, CadoniPitzalis25,CadonideLima26} does not imply that an astrophysical object must gain gravitating mass in that way \citep[c.f.][]{GuarientoHorvath11,GuarientoFontanini12,AfshordiGuariento14,MacielGuariento15}.

Observational searches for signatures of cosmological mass growth $\propto a^\kappa$ in other BH populations have met with mixed results \cite{CrokerNishimura20,CrokerZevin21,Rodriguez23,AndraeElBadry23,GaoLi2023,Ghodla23,LacyEngholm24,MlinarZwitter24,Amendola24,CalzaFrancesco25,Shah25}.
The challenge is primarily the slowness of cosmological effects, i.e. $H_0 \sim 10^{-18}~{\rm Hz}$. 
Searches for differential growth in BH populations become complicated by systematic effects or the need for age proxies, and any lack of horizons removes the lower-bound BH mass $\sim 2M_\odot$ inherited from neutron star (NS) instability. 
Ultimately, known local GR models cannot accommodate the range of mass growth proposed by theory and measured in data.
This prevents the use of precision binary pulsar systems \citep[e.g.][]{WeisbergHuang16,KramerStairs21}, whose orbital dynamics could constrain $\kappa$ for NSs and BHs directly given a one-parameter family of solutions.

Here, we take the first significant step toward this goal.
We demonstrate that GR can locally accommodate cosmologically coupled mass growth by \emph{construction}: we exhibit a well-behaved GR solution that reproduces cosmologically coupled phenomenology at fixed $\kappa$.
We show that Keplerian orbits are affected $\propto a^\kappa$ and use binary pulsar data to bound $\kappa$ for NSs.
We find that $\kappa = 3$ naturally introduces a stressed region with vacuum symmetries, as predicted by unrelated global methods \cite{CrokerWeiner19}. 
Finally, we examine the means by which our solution couples and discuss future directions.
Following observational literature, we define mass via impact on the local gravitational environment.
Following \citet{Gliner66}, we define energized vacuum as a contribution to the stress-tensor with four equal eigenvalues, e.g. a cosmological constant or the non-singular BH interior of \citet{Dymnkova02}, and remain agnostic to possible microphysical origin.
We set the speed of light $c := 1$, adopt the {\tiny$(-,+,+,+)$} signature and use Riemann sign $R^\alpha_{\beta\gamma\delta} := \Gamma^\alpha_{\beta\gamma,\delta} - \Gamma^\alpha_{\beta\delta,\gamma} + \cdots$.

\section{Definition of a cosmologically coupled point model}
It is usually assumed that cosmological background effects cannot impact local systems more strongly than accelerations $O(H^2)$.
For example, construction of the Parameterized Post-Newtonian (PPN) formalism asserts ``\dots~conservation of rest-mass energy \dots'' and that ``\dots the time evolution of the [local] system is governed by the motions of its constituents, [so] we have $\partial/\partial t \sim \vec{v} \cdot \nabla$\dots'' \citep{WillTEGP18}.
These assumptions lead to the conclusion that ``for all solar-system or astrophysical applications \dots we are justified in ignoring the surrounding cosmological setting.'' \citep{WillTEGP18}.
\citet{Kopeikin12} has generalized post-Newtonian methods to $\mathcal{O}\left(H\right)$ effects perturbatively at small scales within a RW background.
  Restriction to stationary sources then allows well-defined notions of ``... conserved mass, dipole moment and other multipoles ... independent of the Hubble parameter.'' \citep[][p.7]{Kopeikin12}
  This construction points directly toward the necessary generalization: a cosmologically coupled local solution cannot be stationary.

The lack of time-translation symmetry in RW cosmologies allows us to construct a solution for specific gravitating sources where mass is not conserved.
We assert a flat $\Lambda$CDM RW cosmology, in conformal RW coordinates and framing, and add to it a symmetric piece $h_{\sigma\nu}$
\begin{align}
  \d s^2 := a(\eta)^2\left[\eta_{\sigma\nu} + h_{\sigma\nu}\right]\d x^\sigma\otimes\d x^\nu. \label{eqn:standard_metric}
\end{align}
The existence of the RW cosmology and its symmetries enables a Scalar-Vector-Tensor decomposition (SVT) of $h_{\sigma\nu}$ \cite{York74}.
Let $r$ be comoving radius and define,
\begin{align}
  h_{\sigma\nu} := 2\delta_{\sigma\nu}\frac{GM}{ar}\left\{a^\kappa - |\kappa|Qr\left[\ln\left(\frac{r}{r_0}\right) - 1\right]\right\}. \label{eqn:beautiful_toy}
\end{align}
Because \cref{eqn:beautiful_toy} $\propto \delta_{\sigma\nu}$, the V and T portions of $h_{\sigma\nu}$ vanish by construction leaving only S DOFs \citep[][inspect Eqs.~(10-13)]{HuAmazingNotes04}.
The constant $\kappa$ is astrophysical: the observationally constrained cosmological coupling strength \cite{CrokerNishimura20,CrokerZevin21,Farrah23a,Farrah23b,Rodriguez23,AndraeElBadry23,GaoLi2023,Ghodla23,LacyEngholm24,MlinarZwitter24,Amendola24,CalzaFrancesco25,Farrah25,Shah25}.
The constant $Q \sim H_0$ is determined from choice of causal domain in $(a,\kappa)$ and will be set to zero at $r > r_0$ during junctioning to satisfy SVT radial decay hypotheses \cite{York74}.
The first term of \cref{eqn:beautiful_toy} reduces to Eqn.~(54) of \citet[][]{Kopeikin12} when $\kappa = Q = 0$.

We build a model by linearizing Einstein's equations in $h_{\sigma\nu}$.
This guarantees that gauge-invariant \citet{Bardeen1980} scalar potentials $\Phi_A$ and $\Phi_H$ exist and encode the entire S content of \cref{eqn:beautiful_toy}.
Furthermore V and T DOFs remain absent.
The physical DOFs are \emph{specified} by \cref{eqn:beautiful_toy} and there are no aphysical DOFs.
We now demonstrate that the family of solutions \cref{eqn:standard_metric,eqn:beautiful_toy} is causal in linearized GR and geometrically encodes cosmologically coupled mass growth for local gravitating systems.

\section{Cosmological coupling is not a coordinate artifact}
We demonstrate that $\kappa$ is physical via the spectra of the Weyl and stress-tensor linear operators (endomorphisms).
Any $\kappa$ dependence in these diffeomorphism invariant quantities encodes physical content of the solution.
The Weyl portion $C^\sigma_{\pi\gamma\delta}$ of the Riemann tensor can be non-zero in vacuum and thus describes the free gravitational field \cite[e.g.][]{Hawking66}.
Weyl shares the symmetries of Riemann, so raising an index produces an operator on the 6-dimensional space of determinant-convention bivectors, 
\begin{align}
   \frac{1}{4}\tensor{C}{^{\sigma\pi}_{\gamma\delta}}(\partial_\sigma\wedge\partial_\pi) \otimes (\d x^\gamma \wedge \d x^\delta).
\end{align}
The operator is diagonal in $\{\partial_\sigma \wedge \partial_\pi\}$ with eigenvalues
\begin{align}
  \mu_{\eta\wedge r} = \mu_{\theta\wedge\beta} = -\frac{2GM}{(ar)^3}\left(a^\kappa + \frac{2|\kappa|Qr}{3}\right) \label{eqn:gravity_eigen_1} \\ 
  \mu_{\eta\wedge \theta} = \mu_{\eta\wedge\beta} = \mu_{r \wedge \theta} = \mu_{r \wedge \beta} = -\frac{\mu_{\eta\wedge r}}{2}, \label{eqn:gravity_eigen_2}
\end{align}
where $\theta$ and $\beta$ are polar and azimuthal angles, respectively.
Note that \cref{eqn:gravity_eigen_1,eqn:gravity_eigen_2} agree with the Schwarzschild solution's Weyl spectrum when $\kappa = 0$ and $a$ is fixed.
By inspection, all eigenvalues depend on $\kappa$ and so the coupling strength is unambiguously a physical property of the free gravitational field.

To examine the stress, we compute the Einstein operator $G^\mu_\nu \propto T^\mu_\nu$. 
Introduce dimensionless quantities,
\begin{align}
  \epsilon := \frac{GMa^\kappa}{ar} \qquad \alpha := Har \qquad q := Q/Ha \label{eqn:dimensionless_combos}
\end{align}
with $q(a) \sim \mathcal{O}(10)$.
We expect stellar gravitational collapse for $z \leqslant 30$, during which $\alpha \sim 10^{-15}\,(r/{\rm AU})$ and $\epsilon \sim 10^{-8}(1+z)^{-\kappa}\,(M/M_\odot)\cdot(r/{\rm AU})^{-1}$.
Let $P := w \rho$ and $\rho = 3H^2/8\pi G$ be the cosmological background pressure and energy density, respectively.
Now define
\begin{align}
    \gamma := 3(w + 1) + \kappa(3w - 1) - 2\kappa^2.
\end{align}
With terms sorted by decreasing magnitude below the horizon $\alpha < 1$, we find $G^\theta_\theta = G^\beta_\beta = G^r_r$ and
\begin{align}
  G^\eta_\eta &= - 8\pi GMa^\kappa\delta^3(r) -\frac{2\alpha\epsilon|\kappa|q}{a^\kappa(ar)^2} - \frac{3\alpha^2}{(ar)^2} - \frac{6\alpha^2\epsilon\kappa}{(ar)^2}  \label{eqn:Getaeta} \\
  G^r_r &= \frac{3\alpha^2w}{(ar)^2} + \frac{\alpha^2\epsilon\gamma}{(ar)^2} - \frac{3(1+w)}{(ar)^2}\frac{\alpha^3\epsilon|\kappa|q}{a^\kappa}\ln\left(\frac{r}{r_0e}\right) \label{eqn:Grr} \\
  G^\eta_r &= -\frac{2\alpha\epsilon\kappa}{(ar)^2}, \label{eqn:Getar}
\end{align}
where the Dirac $\delta^3(r)$ is recovered in a limiting analysis (see Appendix \ref{sec:dirac_delta}).
Covariant conservation of stress-energy $\nabla_\mu T^\mu_\nu = 0$ follows trivially from the contracted Bianchi identities.
Eigenvalues follow from the characteristic equation,
\begin{align}
  \left[\left(G^\eta_\eta - \lambda\right)\left(G^r_r - \lambda\right) + \left(G^r_\eta\right)^2\right]    \prod_{\{\theta, \beta\}}\left(G^{\rm i}_{\rm i} - \lambda\right) \label{eqn:characteristic}
  = 0,
\end{align}
where the $G^r_\eta$ term enters positively due to skew-symmetry.
By inspection, all eigenvalues depend on $\kappa$, and so the coupling strength is unambiguously also a physical property of the matter sector.

\section{The linearized cosmologically coupled point mass satisfies the Dominant Energy Condition (DEC)}
In order for the linearized model of a cosmologically coupled point mass to be acceptable, all causal observers must perceive causal flow of stress-energy in regions of astrophysical relevance.
This is codified by the DEC and expressed as constraints on the eigenvalues.
Let $\lambda_-$ be the timelike eigenvalue and $\lambda_{\rm s}$ be any spacelike eigenvalue.
Then $\lambda_- \leqslant 0$ and $|\lambda_{\rm s}/\lambda_-| \leqslant 1$ is sufficient to satisfy the DEC.

The first step is to guarantee eigenvalues $\lambda \in \mathbb{R}$.
The system has $\lambda \in \mathbb{R}$ if and only if the discriminant of the quadratic roots in \cref{eqn:characteristic} remains non-negative,
\begin{align}
  \left(G^\eta_\eta + G^r_r\right)^2 - 4\left[\left(G^\eta_r\right)^2 + G^\eta_\eta G^r_r\right] \geqslant 0. \label{eqn:discriminant}
\end{align}
We begin with $q=0$ to understand the role of $q\neq0$.
For $q=0$ and $\alpha \ll 1$, \cref{eqn:discriminant} is satisfied if and only if,
\begin{align}
  ar \geqslant \sqrt{\frac{4GM|\kappa| a^\kappa}{3Hc(1+w)}}
  \label{eqn:threshold}
\end{align}
To develop some intuition, we note that the first observational evidence for coupled mass growth was in galactic SMBHs \cite{Farrah23a,Farrah23b, Farrah25}.
  The Milky Way central SMBH Sgr A$^*$ has mass $4.3 \times 10^{6}\,M_\odot$, so \cref{eqn:threshold} gives a distance $\sim 10^2~\rm pc$ today.
  But this number is conservative because physically realistic environments of coupled objects contribute%
  \footnote{In linearized GR, $\rho_{\rm env}$ adds to \cref{eqn:standard_metric} with the typical Poisson contribution $\propto \int_{\vec{x}'} \rho_{\rm env}(\vec{x}')~\d \vec{x}'/|\vec{x}-\vec{x}'|$, in proper distances.}
  additional energy density $\rho_{\rm env}$ to $G^\eta_\eta$, but negligibly%
  \footnote{Peculiar flux $\propto \rho_{\rm env}Har$ enters $G^\eta_r$, but is always dominated by $\rho_{\rm env}$ in $G^\eta_\eta$.}
  to $G^r_r$ and $G^\eta_r$.
  Fix $a=1$ and note the $G^r_r$ contribution to \cref{eqn:discriminant} can be neglected near the object so that,
  \begin{align}
    |G^\eta_\eta| \geqslant 2|G^\eta_r| \implies r \geqslant \sqrt{\frac{\kappa H_0M}{2\pi c\rho_{\rm env}}}.
  \end{align}
  Sgr A$^*$ is immersed within a CDM halo well-described by a Navarro-Frenk-White (NFW) profile
\begin{align}
  \rho_{\rm env}(r) := \frac{\rho_{0\rm h}r_{\rm h}^3}{r\left(r_{\rm h} + r\right)^2},
\end{align}
with mean $r_{\rm h} = 19\,\rm{kpc}$ and $\rho_{0\rm h} = 0.01\,M_\odot/\rm pc^{3}$ \cite{McMillan17}.
Accounting for this environment, we find that a $\kappa = 3$ solution satisfies the DEC at $r > 5.9$ Schwarzschild radii (because $G^\eta_r \propto 1/r^2$ but inner NFW $\propto 1/r$).
For Sgr A$^*$, this is $0.5\,\rm AU$, $24\times$ smaller than the pericenter of S301 \cite{Piran26}, far below where anyone would use linearized GR.

Motivated by this behavior, we now determine how much ``halo'' a coupled point requires, quantified by $Q$, to satisfy the DEC down to its formal Schwarzschild radius.
  For simplicity, \cref{eqn:beautiful_toy} considers a truncated singular isothermal sphere (SIS) \cite[][\S4.3]{BinneyTremaine}, the simplest galactic halo model.
  This halo construction is not unique, but isolates the phenomenology of the coupled point efficiently because its derivatives are succinct and the $1/r^2$ SIS density profile radially matches $G^\eta_r$.
The discriminant \cref{eqn:discriminant} remains non-negative if
\begin{align}
  3\alpha(1+w) \geqslant \epsilon|\kappa|\left(4 - 2qa^{-\kappa}\right).
\end{align}
All terms appearing in this expression are positive for $w$ satisfying the DEC.
So, this inequality is trivially satisfied for $q > 2a^\kappa$, or
\begin{align}
  Q \geqslant \max_{a,\kappa}~2a^{\kappa + 1}H \label{eqn:Q_guidance}
\end{align}
gives $\lambda \in \mathbb{R}$ for all $r$ below \cref{eqn:threshold}.
The range of $\kappa$ relevant to compact objects~\cite{CrokerWeiner19,CrokerRunburg20,CrokerWeiner22} is $-0.2 \lesssim \kappa \lesssim 3$.
The temporal domain relevant to formation by stellar collapse is $a > 1/(1+z)$ with $z \lesssim 30$.
On this physically motivated domain, $Q \geqslant 32H_0$ comfortably suffices to guarantee $\lambda \in \mathbb{R}$, provided the $w \to -1$ limit is avoided.
From inspection of \cref{eqn:Getaeta,eqn:Grr}, $G^\eta_\eta + G^r_r \leqslant 0$ in this near regime and the temporal-radial eigenvalue pair takes the form
\begin{align}
  \frac{2\lambda_\pm}{|G^\eta_\eta + G^r_r|} = -1 \pm \sqrt{1 - \frac{4\left(G^\eta_r\right)^2 + 4G^\eta_\eta G^r_r}{\left(G^\eta_\eta + G^r_r\right)^2}}.
\end{align}
Thus, $\lambda_- < 0$ always.
Because $|\lambda_+| \leqslant |\lambda_-|$, the DEC is always satisfied for the radial direction when $\lambda \in \mathbb{R}$.
Noting that $|G^\eta_\eta| > |G^r_r|$, the angular directions also satisfy the DEC in this regime.

At distances $\alpha \gg \epsilon$, \cref{eqn:threshold} can be satisfied as long as $w$ is not identically de-Sitter.
The $Q=0$ solution is admissible, the eigenvalues simplify to
\begin{align}
  \lambda_- &= -\rho\left(1 + 2\kappa\epsilon\right) + \mathcal{O}\left(\frac{\epsilon}{\alpha}\right) \label{eqn:beautiful_timelike} \\
  \lambda_{+\theta\beta} &= \rho\left(w + \frac{\epsilon\gamma}{3}\right) + \mathcal{O}\left(\frac{\epsilon}{\alpha}\right) \label{eqn:beautiful_spacelike}
\end{align}
and the DEC is satisfied whenever
\begin{align}
  \left|\frac{-\lambda_{+\theta\beta}}{\lambda_-}\right| = \left|w + \epsilon\left(\frac{\gamma}{3} - 2\kappa w\right)\right| \leqslant 1.
\end{align}
This is always true except far into DE domination.

Finally, to produce a solution valid for all spacetime, pick any $Q$ satisfying \cref{eqn:Q_guidance} and choose a radius $r_0$ that satisfies \cref{eqn:threshold} divided by $a$.
An interior solution with $Q \neq 0$ at $r \leqslant r_0$ satisfies the Israel-Darmois junction conditions \cite[][\S3.7]{RelativistToolkit} with an exterior solution built as follows: superpose within $h_{\sigma\nu}$ the $Q=0$ solution and an additional $\kappa = 0$ solution with $M_{r>r_0} := MQ|\kappa|r_0$, where $\kappa$, $M$, and $Q$ are taken from the $r < r_0$ solution.
The $\kappa = 0$ addition encodes the impact of the truncated SIS at $r > r_0$ (see Appendix \ref{sec:junction}).

\section{Curvature contributions $\propto H^2$ need not characterize cosmological impacts on local observables}
\cref{eqn:standard_metric,eqn:beautiful_toy} are a spherically symmetric solution to the linearized Einstein equations for the source \cref{eqn:Getaeta,eqn:Grr,eqn:Getar}.
  The impact on local gravitating systems follows from the invariant Hernandez-Misner-Sharp mass \cite{Hayward96} (through $\mathcal{O}\left(\alpha^2\right)$, see Appendix \ref{sec:hms_mass}).
  For $r < r_0$, we find
  \begin{align}
    E = M\left[a^\kappa + \alpha|\kappa|q + \alpha^2a^\kappa\left(\kappa + \frac{3}{2}\right)\right] + \frac{ar\alpha^2}{2G}. \label{eqn:HMS_mass}
  \end{align}
  The leading order term is a cosmologically coupled mass with growth $\propto a^\kappa$.
  Via Friedmann, the final term is $4\pi(ar)^3\rho/3$, which is the physical cosmological background mass enclosed by a 3-sphere of radius $ar$.
  The remaining terms are integrations over the SIS and a ``cloud'' $\propto 1/r$ both visible in \cref{eqn:Getaeta,eqn:Grr}, and both of which we now show to be observationally irrelevant.

Consider a test timelike observer with worldline tangent field $U : \mathbb{R} \to T\mathcal{M}$, initially at rest so that $U(\eta_0) = U^0 \partial_0$.
Using $g(U,U):=-1$ to resolve $U^0$, the geodesic equation for the radial acceleration becomes 
\begin{align}
  \frac{\d U^r}{\d \tau}\bigg|_{\eta_0} & = -\frac{1}{a}\frac{G M a^\kappa}{(ar)^2}\left[1 + \frac{q |\kappa|\alpha(r)}{a^\kappa}\right], \label{eqn:Newton}
\end{align}
where $\tau$ is the proper time.
Conversion to physical acceleration removes the $1/a$ factor and introduces terms $\mathcal{O}(\alpha^2)$.
We recover Newton's Law of Gravitation sourced by a cosmologically coupled point mass at the origin to a fractional correction from the SIS $\mathcal{O}(\alpha) \sim 10^{-15}$ today, as expected from \cref{eqn:HMS_mass}.
Now hold $a = 1$ to evaluate the precession induced by the SIS today \cite[][Eq.~50]{AdkinsMcDonnell07},
\begin{align}
  \Delta\theta_P({\rm SIS}) = -\frac{2\pi R}{\varepsilon^2}H_0|\kappa|q\left[\sqrt{1-\varepsilon^2} - \left(1-\varepsilon^2\right)\right],
\end{align}
where $R$ is the semi-major axis, $\varepsilon$ (not $\epsilon$) is eccentricity, and we have used standard Keplerian relations between axes and semi-latus rectum \cite[e.g.][\S15]{LandauMechanics}.
The uncertainty in the measured precession of the Hulse-Taylor binary pulsar is $4\times 10^{-6}\,{\rm deg}\cdot{\rm yr^{-1}}$ \cite[][Table~2]{WeisbergHuang16}.
We find $\Delta\theta_P({\rm SIS}) \sim -2.5\times 10^{-10}\,{\rm deg}\cdot{\rm yr^{-1}}$ for $q=32$ and $\kappa=3$.
Thus, the SIS is unobservable by 4 orders of magnitude.

For tight binary dynamics, the only observable evidence of coupling is $a^\kappa$ growth.
Although a comprehensive analysis is beyond the scope of this study, an order of magnitude analysis follows from \citet{CrokerNishimura20}.
For a binary of identically coupled objects, conservation of total angular momentum%
\footnote{Conservation of specific angular momentum, as would be expected with material accretion (c.f. \S\ref{sec:not_accretion}), alters \cref{eqn:Pbdot} by an $O(1)$ constant scaling \citep[e.g.][]{Hadjidemetriou63,DamourTaylor91}.} 
gives semi-major axis evolution $R \propto a^{-3\kappa}$ and reduced mass $\mu \propto a^\kappa$.
So, Kepler's third law gives an orbital period $P_{\rm b} \propto a^{-5\kappa}$ and
\begin{align}
  \dot{P}_{\rm b} = -5\kappa_{\rm NS}H_0P_{\rm b}. \label{eqn:Pbdot}
\end{align}
The double pulsar PSR J0737–3039A/B gives stronger constraints here, with $\dot{P}_{\rm b} = -1.247920(78)\times 10^{-12}$ in agreement with GR radiative loss calculations that assume $\kappa_{\rm NS} := 0$ \cite[][Table~IV]{KramerStairs21}.
We find that $|\kappa_{\rm NS}| \lesssim 8\times 10^{-4}$, which is $\sim 10^2\times$ smaller than $\kappa_{\rm TOV} \sim -0.07$ from naive application of the \texttt{SLy} NS equation of state \cite{DouchinHaensel01} with Tolman-Oppenheimer-Volkoff following \citet{CrokerWeiner19}.
If that prescription is correct \citep[c.f.][]{Mistele23}, NS structure detaches the stress from the cosmology like the electromagnetic cavity of \citet{Kopeikin15} or NSs contain equilibrated energized vacuum cores \cite[e.g.][]{AraujoLima24, AraujoLugones25, JampolskiRezzolla26} that contribute $\kappa_{\rm core} > 0$ such that $\kappa_{\rm NS} = \kappa_{\rm TOV} + \kappa_{\rm core} \sim 0$.

\section{Mass growth $\propto a^3$ implies a region with energized vacuum, as predicted by global analyses}
The entirely background-order analysis of \cite{CrokerWeiner19} concludes that $\kappa = 3$ can be expected for cosmologically coupled non-singular BHs, which contain energized vacuum interiors.
For simplicity, consider the epoch of matter domination, where $w \sim 0$.
Substituting these values of $\kappa$ and $w$ into the far eigenvalues at $q = 0$, \cref{eqn:beautiful_timelike,eqn:beautiful_spacelike} give
\begin{align}
  \lambda_- &= -\rho - 6\rho\epsilon(r) \qquad \lambda_{+\theta\beta} = -6\rho\epsilon(r),
\end{align}
where $\epsilon(r)$ is defined in \cref{eqn:dimensionless_combos}.
This is the superposition of the cosmological dust background and a spherical cloud of energized vacuum with radial profile $1/ar$.
In matter domination $\rho \propto 1/a^3$ and so the cloud maintains a constant energy density in time.
Already in linearized GR, a coupled point mass with $\kappa=3$ is accompanied by a source of energized vacuum, as predicted by the global analysis of \citet{CrokerWeiner19}.

\section{The coupled point mass does not grow via accretion}
\label{sec:not_accretion}
From the perspective of a comoving observer, the cosmological background and the SIS remain at rest relative to the point mass at the origin.
Most importantly, neither the cosmological background energy density (explicitly) nor the choice of $Q$ impact the rate of mass growth of the coupled point $\propto \kappa H_0 M$ today.
To better understand how the coupled point mass grows, write covariant conservation of stress-energy in the RW $\{\partial_\mu\}$ framing for the timelike 4-current $\tensor{T}{^\mu_0}$,
\begin{align} 
-\nabla_\mu \tensor{T}{^\mu_0} = \partial_0\left(\rho + \rho_{\rm c}\right) - \partial_j \tensor{T}{^j_0} - F_0 = 0. \label{eqn:cons}
\end{align}
We have abbreviated the gravitational energy and momentum exchange terms \cite[][\S5.3]{WeinbergGR},
\begin{align}
    F_0 := \tensor{\Gamma}{^\mu_{\mu\rho}} \tensor{T}{^\rho_0} - \tensor{\Gamma}{^\rho_{\mu 0}}\tensor{T}{^\mu_\rho}. \label{eqn:defn_F0}
\end{align}
Here $\rho_{\rm c}$ are the non-background contributions to \cref{eqn:Getaeta}.
The spatial 3-current divergence
\begin{align}
  \partial_j \tensor{T}{^j_0} = -\frac{\kappa MHa^{\kappa + 1}}{2\pi (ar)^3} \neq 0, 
\end{align}
is typically interpreted as quantifying the movement of material into or out of a region, i.e. accretion.
But if we compute the gravitational exchange from \cref{eqn:defn_F0} and substitute it into \cref{eqn:cons}, we find that
\begin{align} 
  \partial_0\rho_{\rm c} &= \partial_j\tensor{T}{^j_0} + \left\{\frac{3M|\kappa|}{4\pi}\left[\frac{H^3a^\kappa\omega}{r\,{\rm sgn}\,\kappa} - \frac{HQ}{(ar)^2}\right] - \partial_j\tensor{T}{^j_0}\right\}, \label{eqn:cloud_change}
\end{align}
where $\omega := \kappa - 3w-4$ for concision and we have removed the background conservation relation for clarity.
Any flow of material is exactly canceled by a portion of $F_0$. 
Furthermore, the remaining terms from $F_0$ match the radial profiles of the fourth and second terms in $G^\eta_\eta$ (everything else is point-mass or background).
So, the shape of the material cloud never changes.
Now integrate \cref{eqn:cloud_change} over a 3-sphere of radius $r_0$ set by \cref{eqn:threshold} to get the rate-of-change for the enclosed mass, excluding the coupled point.
Because $Q \sim H_0$ by construction, the integrated quantities become $\mathcal{O}\left(H_0^2\right)$ and $\mathcal{O}\left(H_0^{3/2}\right)$.
The change in enclosed cloud mass fails to feed the point mass' growth by $\sqrt{H_0}$.

\section{Discussion}
  We have shown in detail that cosmologically coupled phenomenology with flexible $\kappa$ can exist in linearized GR.
  We emphasize that the existence of a viable linearized GR model does not establish that cosmological coupling must occur.
As with any GR solution, data will determine which scenarios ``on paper'' are most appropriate for describing nature.
  Although we have used the SVT machinery often employed in PT to guarantee that only scalar DOFs are present and well-defined, \cref{eqn:Getaeta,eqn:Getar,eqn:Grr} do not contain signed fluctuations about the cosmological background.
  A next question is whether coupled phenomenology can be recovered from the spatial evolution of first-order PT DOFs \citep[e.g.][]{Bardeen1980,KodamaSasaki, MaBertschinger95, HamazakiKodama96, KopeikinPetrov14, PetrovKopeikinBook}.
  The Lagrangian perturbative treatment of \citet{KopeikinPetrov14} (see also \cite{PetrovKopeikinBook}) is particularly well-suited for such studies, as the formalism is explicitly constructed for local tests in genuine cosmological settings.
Whether and how alternative formulations of gravitation \cite[e.g.][]{LasenbyDoran98,GeometricAlgebraBook} could manifest cosmological coupling remains unexplored.

Another natural question is how can \cref{eqn:standard_metric,eqn:beautiful_toy} accommodate a wide range of $\kappa$ while \citet{CadoniSanna23} conclude that $\kappa = 1$?
While a detailed analysis is beyond the scope of this work, the reason is essentially that the metric ans\"{a}tze are physically inequivalent.
The conformally scaled ansatz given in \citet[][Eqn.~(2.2)]{CadoniSanna23} has a time-independent angular sector.
In spherically symmetric \emph{static} spacetimes, any modulation of the angular sector from gravity can be absorbed into a redefinition of the radial coordinate.
Because our \cref{eqn:beautiful_toy} contains a time-dependent angular sector, however, redefinition of the radial coordinate introduces off-diagonal flux terms $\propto \partial_\eta h_{00}$.
For general $\kappa$, these fluxes cannot be removed without making the scale factor position dependent.
Only when $\kappa = 1$ and $Q=0$ does \cref{eqn:beautiful_toy} become time-independent, in which case \cref{eqn:standard_metric,eqn:beautiful_toy} can be brought to the form of \citet{CadoniSanna23}.

Our result is ``one-sided'' in that it does not establish that the cosmology must respond to the local object's stress.
Investigations in this direction could proceed by generalizing to compact RW spacetimes (e.g. closed RW, \cite{NandraLasenby12}).
In this setting, a single object is sufficient to source the RW background.
We caution that \cref{eqn:beautiful_toy} should not be used in Einstein's equations beyond the linear regime because the solution ansatz itself is constructed only through linear order.

\section{Summary}
Cosmologically coupled objects with passive mass evolution $\propto a^\kappa$ have been shown to possess phenomenologically interesting consequences, in particular providing a source for accelerated late-time expansion with observationally relevant time-dependence when the mass growth parameter $\kappa \sim 3$.
We have constructed a simple GR model for a cosmologically coupled point with invariant Hernandez-Misner-Sharp gravitating mass $\propto a^\kappa$.
For astrophysically relevant values of $\kappa$ and $a$, the solution is causal (as characterized by the DEC) within the regime of validity of the linear approximation.
Our results are unaffected by previous arguments that $\kappa \sim 3$ is not supposed to occur in GR~\cite{Mistele23, WangWangARXIV, Avelino23, Parnovsky24, GaurVisser24} and highlight that cosmological coupling is absent from many GR analyses by construction, not by necessity.
The solution does not obviously contradict existing observations.
Riemann curvature contributions $\propto H^2$ fail to characterize the impact on observables like Keplerian orbital timing, which can now be used to constrain $\kappa$.
Within a matter-dominated universe, $\kappa = 3$ induces an energized vacuum cloud, complementing a demonstration by \citet{CrokerWeiner19} that localized energized vacuum should gravitate $\propto a^3$.
Solutions with large cosmologically coupled mass growth have now been shown to exist in linearized GR.

\begin{acknowledgements}
  We thank the anonymous referees for incisive criticism that significantly strengthened the results of this Manuscript.
  KC thanks the Perimeter Institute for Theoretical Physics and C.S.~Jackson for their hospitality during the preparation of this manuscript.
  CC acknowledges support from the Beus Center for Cosmic Foundations at Arizona State University.
  NF acknowledges funding from the U.S. Department of Energy Grant No. DE-SC0010008.
  GN and AA acknowledge the support of SECIHTI (grants CBF-2023-2024-162 and CBF-2025-I-2795), DGAPA-PAPIIT (grant IA101825), DAIP-UG and the Instituto Avanzado de Cosmología.
  LT is supported by JSPS under KAKENHI 24K22878 and 26K17136 and by the Royal Society under ICA\textbackslash R2\textbackslash 252140.
  RAW acknowledges support from NASA JWST Interdisciplinary Scientist grants NAG5-12460, NNX14AN10G and 80NSSC18K0200 from GSFC.
  JWR acknowledges funding from U.S. Department of Energy Grant No. DE-SC0016021.
  Research at Perimeter Institute is supported in part by the Government of Canada through the Department of Innovation, Science and Economic Development Canada and by the Province of Ontario through the Ministry of Colleges and Universities.

  Analytic results were verified independently using two computer-algebra systems, \texttt{Maxima}~\cite{maxima,toth2005tensor} and \texttt{Mathematica}~\cite{mathematica}, and confirmed numerically using \texttt{NumPy}~\cite{numpy}.
\end{acknowledgements}

\appendix
\section{Dirac delta}
\label{sec:dirac_delta}
To identify appearances of the Dirac $\delta$ and to verify that they do not corrupt our analysis, we set $Q = 0$ and replace the term $\propto 1/ar$ in \cref{eqn:beautiful_toy} with,
  \begin{align}
    \phi_\sigma &:= -\frac{GMa^\kappa}{ar}\frac{2}{\sqrt{\pi}}\int_0^{ar/\sigma\sqrt{2}}e^{-t^2}\,\d t \\
     &= -\frac{GMa^\kappa}{ar} \text{erf}\left(\frac{ar}{\sigma\sqrt{2}}\right).
  \end{align}
Note that $\lim_{\sigma \to 0^+} \phi_\sigma$ regenerates \cref{eqn:beautiful_toy} with $Q=0$.
With this definition, we find
\begin{align}
  \begin{split}
    -\lambda_- =~&\rho + \frac{e^{-\left(ar/\sigma\sqrt{2}\right)^2}}{\left(\sigma\sqrt{2\pi}\right)^3} Ma^\kappa \\
    &+ G\rho Ma^\kappa \left[\frac{4e^{-\left(ar/\sigma\sqrt{2}\right)^2}}{\sigma\sqrt{2\pi}} + \frac{2\kappa}{ar} \text{erf}\left(\frac{ar}{\sigma\sqrt{2}}\right)\right].
    \end{split}
\end{align}
The first term with $M$ satisfies the nascent Dirac $\delta$ requirement,
\begin{align}
  Ma^\kappa \int \lim_{\sigma\to 0^+} \frac{e^{-\left(ar/\sigma\sqrt{2}\right)^2}}{\left(\sigma\sqrt{2\pi}\right)^3} \sqrt{g_3}~\d^3x = Ma^\kappa,
\end{align}
where $g_3$ is the 3-space metric determinant; this normalization is what fixes $\delta^3(r)$ in \cref{eqn:Getaeta} as the covariant delta, $\int \delta^3(r)\sqrt{g_3}\,\d^3x = 1$.
The second term is non-pathological and disappears in the $\sigma \to 0$ limit.
The third term regenerates the $1/r$ pieces of \cref{eqn:Getaeta,eqn:Grr}.
A similar analysis of $\lambda_{+\beta\theta}$ shows no Dirac $\delta$ appears, consistent with \cref{eqn:Newton} and expectations from active gravitational mass in Minkowski spacetime \cite[][\S8.6,~Ex.~20]{SchutzGR}.

\section{Junction Conditions}
\label{sec:junction}
Fix $M$, $\kappa$, and $r_0$.
We junction an interior solution with $Q\neq0$ to an exterior solution with $Q=0$, built from the superposition of 1) a coupled point with $M$ and $\kappa$, and 2) a decoupled ($\kappa = 0$) point with mass parameter $\overline{M}$ to be determined.
We junction at $r=r_0$, a hypersurface with coordinate normal $\partial_r$.
Using $h_{\sigma\nu}$ from \cref{eqn:beautiful_toy}, define
\begin{align}
  h_{\sigma\nu}(r \geqslant r_0) := \left[h_{\sigma\nu}\big|_{\kappa,M} + h_{\sigma\nu}\big|_{\kappa=0,\overline{M}}\right]_{Q=0}. \label{eqn:exterior_h}
\end{align}
The discontinuity of the induced metric $\hat{g}_{\sigma\nu}$ across the junction becomes,
\begin{align}
  \left[\hat{g}_{\sigma\nu}\right]_{r_0} = \frac{2Ga}{r_0}\left(QM|\kappa|r_0 - \overline{M}\right)\hat{\delta}_{\sigma\nu},
\end{align}
so the first junction condition is satisfied if,
\begin{align}
  \overline{M} = Mr_0|\kappa|Q. \label{eqn:nut}
\end{align}

For any superposition of $h_{\sigma\nu}$ defined in \cref{eqn:beautiful_toy} with common $r_0$, extrinsic curvatures on hypersurfaces orthogonal to $\partial_r$ are,
\begin{align}
  K_{\beta\beta} &= K_{\theta\theta} \sin^2\theta \\
  K_{\theta\theta} &= ar - \sum_i GQ_iM_i|\kappa_i|r\left[\ln\left(\frac{r}{r_0e}\right) + 1\right]\\ 
  K_{\eta\eta} &= -\sum_i \frac{GM_i}{r^2}\left[Q_i|\kappa_i|r + a^{\kappa_i}\right].
\end{align}
At $r_0$, all $Q_i$ terms in $K_{\theta\theta}$ vanish, so $[K_{\theta\theta}]_{r_0} = [K_{\beta\beta}]_{r_0} = 0$.
For the exterior solution \cref{eqn:exterior_h}, continuity of $K_{\eta\eta}$ across $r=r_0$ requires
\begin{align}
  \frac{GM}{r_0^2}\left[Q|\kappa|r_0 + a^\kappa\right] = \frac{GM}{r_0^2}a^\kappa + \frac{G\overline{M}}{r_0^2},
\end{align}
which is also satisfied, given \cref{eqn:nut}.
Thus $[K_{\eta\eta}]_{r_0} = 0$ and all second junction conditions are satisfied.

\section{Hernandez-Misner-Sharp (HMS) Mass}
\label{sec:hms_mass}
Let $\mathcal{R}$ be the areal radius, i.e. the square-root of the factor scaling the symmetry 2-spheres.
From \cref{eqn:standard_metric,eqn:beautiful_toy} it is,
\begin{align}
  \mathcal{R} = ar\left(1+\frac{h_{00}}{2}\right).
\end{align}
\citet[][Eq.4]{Hayward96} provides a succinct definition of the HMS energy $E$.
Reintroducing Newton's $G$, we have
\begin{align}
  E := \frac{\mathcal{R}}{2G}\left(1 - g^{\sigma\nu}\partial_\sigma \mathcal{R}\,\partial_\nu \mathcal{R}\right).
\end{align}
This quantity is a geometrical invariant for any spherically symmetric spacetime, including those non-stationary.
At linear order in $h_{\sigma\nu}$, separating the background RW contribution, then sorting by order and dropping terms $\mathcal{O}(\alpha^3)$, we find \cref{eqn:HMS_mass}.

\bibliographystyle{apsrev4-2}
\bibliography{main}{}

\end{document}